\documentclass[sigconf]{acmart}
\usepackage[ruled,vlined,linesnumbered]{algorithm2e}
\usepackage{enumitem}

\usepackage{epsfig}

\newcommand{\ww}{{\sc WW}}

\newcommand{\WWClass}{}
\newcommand{\wwbacronym}{{Web-Enabled Simulation}}
\hypersetup{bookmarksdepth=-2}

\acmConference[GI 2020]{GI 2020: 8th International Workshop on Genetic Improvement}{October 2020}{Korea}
\acmBooktitle{8th International Workshop on Genetic Improvement}
\setcopyright{none}

\begin{document}

\title{WES: Agent-based User Interaction Simulation on~Real~Infrastructure\vspace{100\in}}
\titlenote{This paper appears in GI 2020: 8th International Workshop on Genetic Improvement.}

\author{John~Ahlgren, Maria~Eugenia~Berezin, Kinga~Bojarczuk, Elena~Dulskyte, Inna~Dvortsova, Johann~George, Natalija~Gucevska, Mark~Harman, Ralf~L\"ammel, Erik~Meijer, Silvia~Sapora, Justin~Spahr-Summers}
\authornote{Author order is alphabetical.\\\mbox{}\ \ Correspondence to Mark Harman ({\tt markharman@fb.com}).}

\affiliation{%
  \institution{FACEBOOK Inc.\vspace{100\in}}
}

\begin{abstract}
We introduce the {\wwbacronym} (WES)   research agenda, and describe FACEBOOK's {\ww} system. We describe the application of {\ww} to reliability, integrity and privacy at FACEBOOK\footnote{`FACEBOOK', refers to the company, while `Facebook' refers to the specific product.}, where it is used to simulate social media interactions  on an infrastructure consisting of hundreds of millions of lines of code. The WES  agenda draws on research from many areas of study, including Search Based Software Engineering, Machine Learning, Programming Languages, Multi Agent Systems, Graph Theory, Game AI, and AI Assisted Game Play. We conclude with a set of open problems and research challenges  to motivate wider investigation.
    \end{abstract}
    
\renewcommand{\shortauthors}{Ahlgren et al.}

\maketitle

\section{Introduction}
A {\wwbacronym} (WES) is a simulation of the behaviour of a community of users on a software platform. 
It uses a (typically web-enabled) software platform to simulate real-user interactions and social behaviour on the real platform  infrastructure, isolated from production users.
Unlike traditional simulation~\cite{kleijnen:supply,michel:multi},
in which a model of reality is created, a WES system is thus built on a real--world software platform.

In order to model users' behaviour on a WES system, a multi agent-based approach is used, in which each agent is essentially a bot that simulates user behaviour. 
This user behaviour could be captured in a rule-based system, or could be learnt, either supervised from examples, or unsupervised in a reinforcement learning setting. 

The development of approaches to tackle the challenges posed by WES may thus draw heavily on recent advances in machine learning for software games, a topic that has witnessed many recent breakthroughs~\cite{oriol:alphastar}.

In this paper we set out the general principles for WES systems. 
We outline the development of FACEBOOK‘s {\ww} simulation of social media user communities, to illustrate  principles of and challenges for the WES research agenda. 
{\ww} is essentially a scaled down simulation of FACEBOOK's platform, the actions and events of which use the same infrastructure code as the real platform itself. 
We also introduce two new approaches to testing and optimising systems: Social Testing and Automated Mechanism Design.

Social Testing tests users' interactions with each other through a platform, while
Automated Mechanism Design combines Search Based Software Engineering (SBSE) and Mechanism Design to automatically find improvements to the platforms it simulates.

Like any software testing system, the WES approach helps find and fix any issues, e.g., with software changes and updates. 
In common with testing systems more generally,  WES operates in a safely isolated environment. 
The primary way in which WES builds on existing testing approaches lies in the way it models behaviour. 
Traditional testing focuses on system behaviour rather than user behaviour, whereas WES focuses on the interactions between users mediated by the system. 

It is a subtle shift of emphasis that raises many technical and research challenges. Software systems involve increasing levels of social interaction, thereby elevating the potential for issues and bugs relating to complex interactions between users and software. 
It is the emergence of these kinds of social bugs and issues that necessitate the need for a WES-style approach, and the research agenda that underpins it.
FACEBOOK’s WW simulation is WES that uses bots that try to break the community standards in a safe isolated environment in order to test and harden the infrastructure that prevents real bad actors from contravening community standards.

\noindent
{\bf Widespread Applicability:}
Community behaviour  is increasingly prevalent in software applications, for example  for travel, accommodation, entertainment, and shopping.
These systems use social interactions so that
each user can benefit from the collective experience of other users. 
Although this paper focuses on FACEBOOK's {\ww} system, the concepts and approach could also find application in platforms used by other organisations.

\noindent
{\bf Realism:}
WES interactions between bots are achieved through the {\em real} platform infrastructure, 
whereas a more  traditional simulation approach would first {\em model} this  infrastructure. 
This is important because the platform infrastructures that mediate user interactions are increasingly complex. 
For instance, FACEBOOK's {\ww} 
 simulation 
is built on a social media infrastructure consisting of several hundreds of millions of lines of code.
While a traditional simulation modelling approach is applicable, there are many issues that are better understood using a WES approach, as we shall see.

\newpage
Platform realism does not necessarily  mean that the interactions between users need to be realistic representations of the end users' experience.
A WES system could, for instance, allow engineers to experiment with new features for which, by definition,  there is no known realistic user behaviour.
It can also be used to focus on atypical behaviours of special interest to engineers, such as those of bad actors.
A WES system could even be used for counter-factual simulation; modelling what users {\em cannot} do.
We use the terms `platform realism' and `end user realism' for clarity.
The term `platform realism' refers to the degree to which the simulation uses the real platform.
The term `end user realism' refers to the degree to which simulated interactions faithfully mimic real users' interactions.
The former is  inherent to the  WES approach, while the latter may be desirable, but is not always essential.

\noindent
{\bf Opportunities for Researchers:}
It may not be possible for researchers to experiment with WES systems directly (for example, where they are built from proprietary software platforms).
Nevertheless, many open  questions can be tackled using traditional simulation, with results extended or extrapolated for application to WES systems.
Researchers can also experiment with and evaluate novel techniques and approaches using WES systems built from open source components.

We report on our plans for the further future development of {\ww}.
The WES  research agenda resides at an exciting intersection between 
many topics including, 
but not limited to,  
Search Based Software Engineering (SBSE)~\cite{mhamyz:acm-surveys}, 
Multi Agent Systems~\cite{wooldridge:agent}, 
Machine Learning~\cite{serrano:grokking}, 
Software Testing~\cite{bertolino:testing},
Graph Theory \cite{west:graph},
Game Theory \cite{myerson:game},
and
Game AI \cite{yannakakis:game}. 
We hope that this paper will serve to stimulate interest in activity in the development of research on WES approaches.

The primary contributions of this paper are:
\begin{enumerate}
    \item The introduction of the WES approach to on-platform simulation of real-world software user communities;
    \item The introduction of the concepts of Automated Mechanism Design and Social Testing, both of which are relevant to WES systems, but also have wider applications;
    \item An outline of the FACEBOOK {\ww} system; an example of a WES system, applied to social media user communities;
    \item A list of open problems and challenges for the WES research agenda.
\end{enumerate}

\section{{\wwbacronym}}
\label{sec:WES}

A WES simulation can be seen as a game, in which we have a set of players that operate to fulfil a certain objective on a software platform. 
This observation connects research on WES systems with research on AI Assisted Game Play~\cite{oriol:alphastar}.
In an AI Assisted Game, reinforcement learning can be used to train a bot to play the game, guided by a reward (such as winning the game or achieving a higher score). 
Similarly, a WES simulation can also use reinforcement learning to train bots.

For example, with FACEBOOK's {\ww} simulation, we train bots to behave like bad actors, guided by rewards that simulate their ability to behave in ways that, in the simulation, contravene our community standards~\cite{community-standards}.
The users whose behaviour is stimulated by other WES approaches could be end-users of the software platform but, more generally, could also be any software user  community. 
For example, a simulation of the users of a continuous integration system, would be a simulation of a developer community, while an App Store WES system may involve both developers and end users.

\noindent
We define the following generic concepts that we believe will be common to many, if not all, WES systems:

\textbf{Bot}: A bot is an autonomous agent. 
Note that bots can create `new' data as they execute.
For instance, social networking bots may weave connections in a social graph as they interact, 
`just as the Jacquard loom weaves flowers and leaves'~\cite{lovelace:sketch}.

\textbf{Action}: An action is a potentially state-changing operation that a bot can perform on the platform.

\textbf{Event}: An event is a platform state change that is visible to some set of users. 

\textbf{Observation}:
An observation of the platform's state  does not change the platform state.
It is useful to distinguish actions (state changing), from observations (which are pure functional). 
This means that some apparent observations need to be decomposed into action and observation pairs.
For example, the apparent observation  `read a message', may update notifications (that the message has been read).
Therefore, it is decomposed into the (state-changing) action of getting the message (which happens once) and the observation of reading the message (which is side--effect free and can occur multiple times for those messages that permit multiple reads).

{\bf Read-only bot}: a read-only bot is one that cannot perform any actions, but can observe state.
Read-only bots can potentially operate on real platform data, because they are side--effect free, by construction.

{\bf Writer bot:} a writer bot that can perform actions and, thereby, may affect the platform state  on which it acts (e.g. the social graph in the case of social media applications).

{\bf Fully isolated bot}: a fully isolated bot can neither read from not write to any part of state that would affect real user experience, by construction of the isolation system in place (See Section~\ref{sec:isol}).

\textbf{Mechanism}: The mechanism is the abstraction layer through which a bot interacts with the platform.
Actions, events and observations may be restricted and/or augmented by the mechanism.
For instance, the mechanism might constrain the actions and events so that the bot can only exhibit a subset of real behaviours of particular  interest, such as rate limiting, or restricted observability.
The mechanism might also augment the available actions and events to explore novel approaches to the platform configuration, products and features,  before these are released to end users.
This opens up the potential for automated mechanism design, as we discuss in Section~\ref{sec:md}.
The mechanism is also one way in which we achieve isolation (see Section~\ref{sec:isol}).

One obvious choice for a mechanism would be to provide a bot with a user interface similar to the one that the GUI offers to a real user. 
However, there is no need for a WES system to interface through the GUI.
There may be practical reasons why it is advantageous to implement interactions more deeply within the infrastructure.
For example, it is likely to be more efficient for a WES bot to bypass some levels of abstraction between a normal user and the core underlying platform.
Of course, there is a natural tension between the twin objectives of  platform realism and efficiency, a topic to which we return in Section~\ref{sec:open}. 

\textbf{Script}:  A script run is akin to a single game episode.
Scripts capture, {\it inter alia}, 
the type of environment created, 
how bots interact,
simulation stopping criteria
and measurement choices.

\textbf{Simulation time}: The simulation  may compress or expand the apparent  time within the simulation in order to simulate a desired use case more efficiently or effectively. 
However, there will always be a fundamental limitation, because the simulation can only be as fast as the 
underlying real platform environment  will permit.
This is another difference between  WES  and  traditional simulation.

\textbf{Monitor}: The monitor captures and records salient aspects of the simulation for subsequent analysis.

\subsection{Bot Training}
Bots behave autonomously, though training to exhibit particular behaviours of interest. 
In the simplest use case, bots merely explore the platform, randomly choosing from a predefined set of actions and observations. 
More intelligent bots use algorithmic decision making and/or  ML models if behaviour. 
The bots could also be modelled to cooperate towards a common goal.

\subsection{Bot Isolation}
\label{sec:isol}
Bots must be suitably isolated from real users to ensure that the simulation, although executed on real platform code, does not lead to unexpected interactions between bots and real users.
This isolation could be achieved by a `sandbox' copy of the platform, or by constraints, e.g., expressed through the mechanism  and/or using the platform's own privacy constraint mechanisms.

Despite this isolation, in some applications bots will need to exhibit high end user realism, which 
poses challenges for the machine learning approaches used to train them.
In other applications where read only bots are applicable, isolation need not necessarily prevent  the bots from reading user information  and reacting to it, but these read only bots cannot take actions (so cannot affect real users). 

Bots also need to be isolated from affecting production monitoring and metrics.
To achieve this aspect of isolation, the underlying platform requires (some limited)  awareness of the distinction between  bots
and real users, so that it can distinguish real user logging data from that accrued by bot execution.
At FACEBOOK, we have well-established gate keepers and configurators  that allow us to do this with minimal intervention on production code.
These gate keepers and configurators essentially denote  a Software Product Line  \cite{clements:spls} 
that could also, itself, be the subject of optimisation  \cite{mhetal:splc14}.

Finally, isolation also requires protection of the reliability of the underlying platform.
We need to reduce the  risk that bots' execution could crash  the production system or affect it by the large scale consumption of computational resources.

\begin{figure}
    \centering
    \includegraphics[width=80mm,scale=0.6]{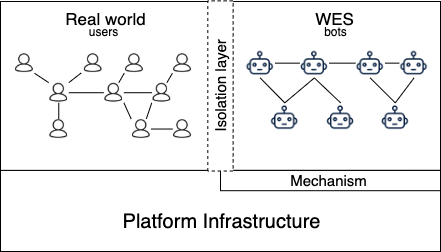}
    \caption{Generic WES System Infrastructure. Real users and bots reside on the same overall platform infrastructure. There is a conceptual isolation layer between them that determines the level of interaction possible, if any, between bots and real users. There is also a mechanism layer that mediates the actions and observations that the bots can perform through the platform.}
    \label{fig:ww_users}
\end{figure}

\subsection{Automated Mechanism Design}
\label{sec:md}

Suppose we want  to experiment with the likely end user behavioural response to new privacy restrictions, before implementing the restrictions on the real platform. 
We could fork the platform and experiment with the forked version. 
However, WES offers a more `light touch' approach: we simply adjust the {\em mechanism} through which bots interact with the  underlying platform in order to model the proposed restrictions.
The mechanism can thus model a possible future version of the platform.

Like all models, the mechanism need not capture all implementation details, thereby 
offering the engineer an agile way to explore such future platform versions.
The engineer can now perform A/B tests through different mechanism parameters,
exploring the differential behaviours of bot communities under different changes.

Using this intermediate `mechanism' layer ameliorates two challenges for automated software improvement  at industrial scales: 
build times and developer acceptance. 
That is, build times can run to minutes~\cite{bell:efficient} or even hours~\cite{hilton:trade}, 
so an automated  search that insists on  changes to the code may incur a heavy computational cost.
Even where cost is manageable, developers may be reluctant to land code changes developed by a machine~\cite{Petke:gisurvey}. 
We have found that a `recommender system' approach sits well with our developers' expectations at FACEBOOK~\cite{ametal:sapfix}.
It ensures that the developer retains ultimate control over what lands onto the code base.

The ease with which the mechanism can be adjusted without needing to land changes into the underlying platform code means that this exploration process can be automated.
Automated Mechanism Design is thus the  search for optimal (or near optimal) mechanisms, according to some fitness criteria of interest. 
In the domain of AI Assisted Game Play, this is akin to changing the rules of the game as the player plays it, for example to make the game more challenging for an experienced player~\cite{kunanusont:n-tuple}.

Borrowing the terminology of economic game theory~\cite{hurwicz:mechanism-design}, we use the term `Automated Mechanism Design'
to characterise the automated (or semi automated) exploration of the search space
 of possible mechanisms through which WES bots interact with the underlying infrastructure.
Automated Mechanism Design is therefore also another application of 
Search Based Software Engineering (SBSE)~\cite{mhamyz:acm-surveys,mhbj:manifesto}.
As with AI Assisted Game Play, we wish to make the platform more challenging, for example, to frustrate bad actors.
However, the applications of Automated Mechanism Design are far wider than this
because it offers a  new approach to automated A/B testing, at volumes never previously considered.

\subsection{Social Testing}
\label{sec:st}
 WES systems bear some relationships to testing, in particular, end-to-end system level testing. 
Indeed, FACEBOOK's {\ww} traces its origins back to observations of the behaviour of multiple executions of FACEBOOK's  Sapienz automated test design platform~\cite{mhetal:ssbse18-keynote}.
However, even with only a single bot, a WES system differs from traditional testing, because a WES bot is trained,
while a traditional test follows a specific sequence of input steps.

Furthermore, unlike end-to-end tests, which typically consider the journey of a {\em single} user through the system and avoid test user  interaction lest it elevate test flakiness~\cite{mhpoh:scam18-keynote}, 
a WES system specifically {\em encourages} test user interaction to model community behaviour. 
Therefore, WES systems can reveal faults that are best explored and debugged at this `community' level of abstraction.
We give several examples of such faults, encountered in our work at FACEBOOK.
Our analysis of the most impactful production bugs indicated that as much as 25\% were social bugs, of which at least 10\% has suitable automated oracles to support a WES approach.

Such social bugs, arising through community interactions, 
require a new approach to testing: {\em Social Testing}; testing emergent properties of a system that manifest when bots interact with one another to simulate real user interactions on the platform.
WES systems are clearly well-suited to Social Testing.
However, we believe other approaches are also viable; Social Testing is an interesting new level of abstraction (lying above system testing levels of abstraction). 
It  is worthy of research investigation in its own right.

In theory, all such `social faults' could be found at the system level.
Indeed, all {\em system} level faults could, {\em in theory}, be found at {\em unit} level.
In practice, however,  it proves necessary to stratify testing.
We believe that social testing (at the whole platform level) is just a new level of abstraction; one that is increasingly important as systems themselves become more social.

\section{FACEBOOK's {\ww}}

At FACEBOOK, we are building a WES system (called
{\ww}), according to the principles set out in Section~\ref{sec:WES}.
{\ww} is an environment and framework for simulating social network community behaviours, with which we investigate emergent social properties of FACEBOOK's platforms.
We are using {\ww} to (semi) automatically explore and find new improvements to strengthen  Reliability, Integrity and Privacy on FACEBOOK platforms.
{\ww} is a WES system that uses techniques from Reinforcement Learning~\cite{sutton:book} to train bots (Multi Agent Reinforcement Learning) and Search Based Software Engineering~\cite{mhamyz:acm-surveys} to search the product design space for mechanism optimisations: Mechanism Design.

Bots are represented by test users that perform different actions on real FACEBOOK infrastructures.
In our current implementation, actions  execute only the back-end code: 
bots do not generate HTTP requests, nor do they interact with the GUI of any platform surface; 
we use direct calls 
to internal FACEBOOK product libraries. 
These users are isolated from production using privacy constraints and a well-defined mechanism of actions and observations through which the bots access the underlying platform code.
However, when one {\ww} bot interacts with another (e.g., by sending a  friend request or  message) it uses  the production back-end code stack, components and systems to do so, thereby ensuring  platform realism.

\subsection{Training {\ww} bots}
\begin{figure}
    \centering
    \includegraphics[scale=0.8]{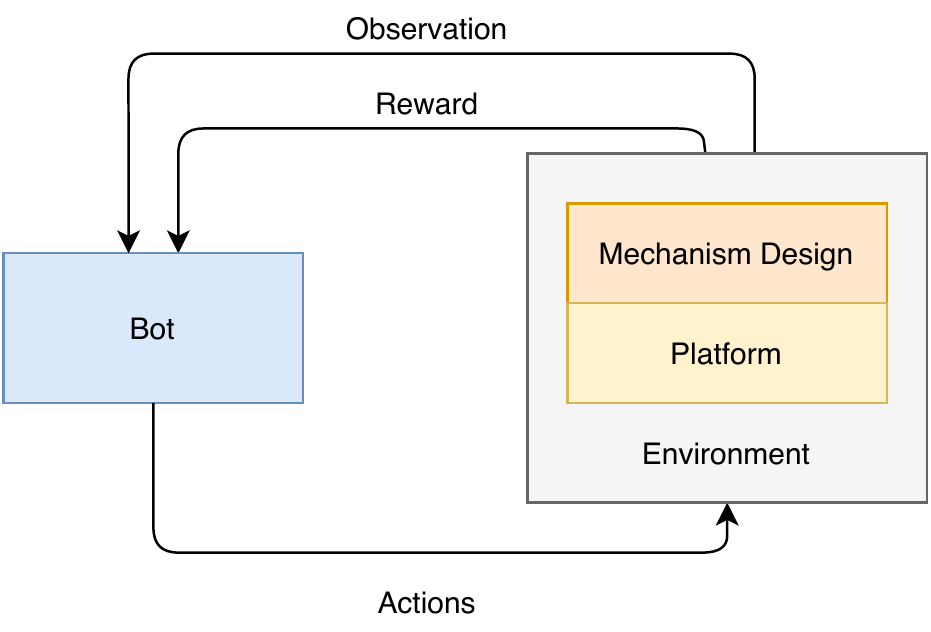}
    \caption{{\ww} Reinforcement Learning Architecture:  bots execute actions in the environment, which in turn, returns an observation, and an eventual reward.
    The platform's code is unchanged by simulation, while the mechanism through which the bots interact with the platform is subject to change  during the simulation process (to explore optimisations of the underlying platform).}
    \label{fig:ww-framework}
\end{figure}

To train bots to simulate real user behaviour,
we use Reinforcement Learning (RL) techniques~\cite{sutton:book}.
Our bot training top level approach is depicted in Figure~\ref{fig:ww-framework}.
As can be seen from Figure~\ref{fig:ww-framework} the  {\ww} bot training closely models that of a typical RL system~\cite{sutton:book}.
That is, a bot executes an action in the environment, which in turn, returns an observation (or current state), and an eventual reward to the bot.
Using this information, the bot decides to take an action, and thus the SARSA (State-Action-Reward-State-Action) loop is executed during a simulation.

However, when considering the environment, we explicitly tease apart the mechanism from the underlying platform.
The platform is out of {\ww}'s control: its code can change, since it is under continual development by developers, but {\ww} cannot change the platform code itself.
Furthermore, {\ww} cannot determine the behaviour of the platform. 
The platform may choose to terminate and/or it may choose to allocate different resources on each episode of the 
simulation.
Furthermore, the social graph at the heart of the database is also continually changing.

The mechanism helps to maintain a consistent interface for {\ww} bots, so that their code is
insulated from such changes. 
It also mediates the actions and observations a bot can make and witness, so that many different mechanisms can be explored without any need to change the platform.
As can be seen from Figure~\ref{fig:ww-framework}, the mechanism is separated from the platform.
Each bot contains its own set of mechanism parameters, so that each can report the fitness of a different mechanism parameter choice.
At the same time, the bots seek to achieve their goals, guided by reinforcement learning.

For example, to simulate scammers and their targets, we need at least two bots, one to simulate the scammer and another to simulate the potential target of the scam. 
The precise technical details of how we impede scammers on {\ww} are obviously  sensitive, so we cannot reveal them here. Nevertheless, we are able to outline the technical principles.

The reinforcement learning reward for the scammer bot is based on its ability to find suitable candidate targets, while the candidate targets are simulated by rule-based bots that exhibit likely target  behaviours. 
The mechanism parameters residing in the scammer bots are used to compute fitness in terms of the mechanism's ability to
impede the scammers in their goal of finding suitable targets.

This use case need not involve a large number of bots.
However, the ability to simulate at scale gives us automated parallelisation for increased scalability of the underlying search, and also supports averaging fitness and reward results over multiple parallel executions.
Such averaging can be useful for statistical testing since results are inherently stochastic.

\subsection{Top Level Implementation}
\begin{figure}
    \centering
    \includegraphics[scale=0.75]{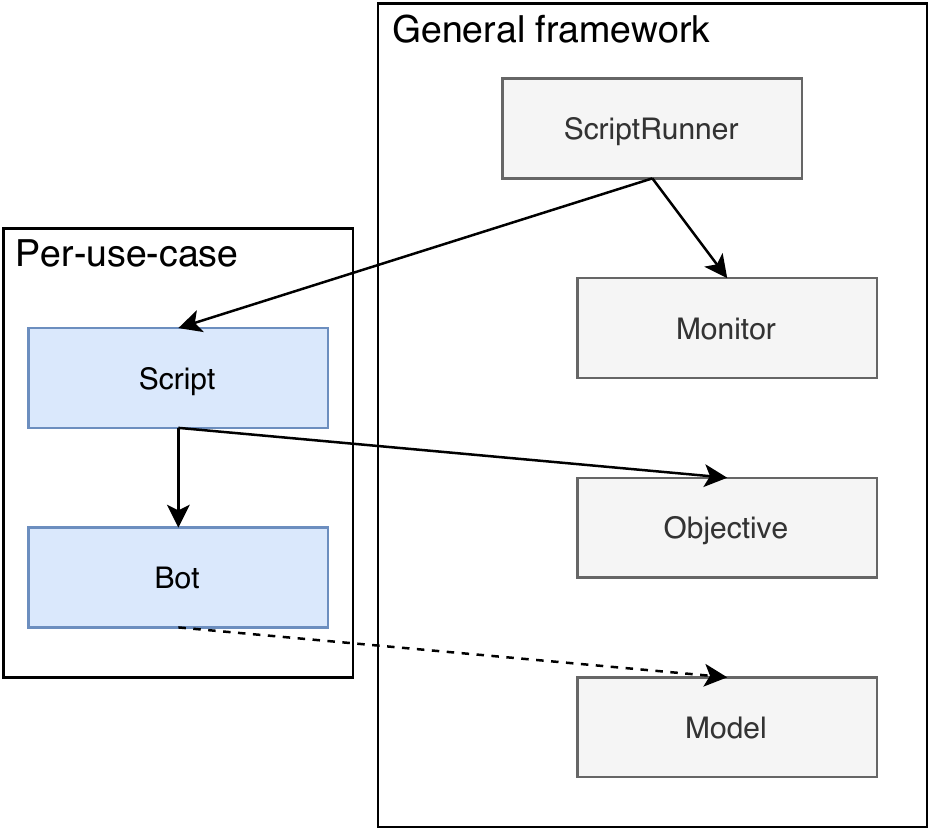}
    \caption{{\ww} implementation:  general framework components  form the core platform, while per-use-case components are specialised to each use case. }
    \label{fig:ww-architecture}
\end{figure}

The top level components of the FACEBOOK {\ww} system  are depicted in Figure~\ref{fig:ww-architecture}.
This is a very high level view; there is, of course, a lot more detail not shown here.
We focus on key components to illustrate overall principles.
{\ww} consists of  two overall subsystems: the general framework classes, which are the core of our simulation platform (and remain unchanged for each use case), and the per-use-case classes (that are tailored for each use case).

\begin{algorithm}[t]
\textbf{Setup}\\
\indent ~~~{\WWClass}Script creates the environment and the {\WWClass}Bots.\\
 \While{{\WWClass}Objective is not reached}{
\indent ~~~Advance the virtual-time clock.\\
\indent ~~~Execute the action of the next {\WWClass}Bot.\\
\indent ~~~Observe and log events and data.\\
 }
\textbf{End}\\
\indent ~~~Finalize the simulation (cleanup).
\caption{Pseudo-code of the {\WWClass}ScriptRunner loop}
\end{algorithm}

\subsubsection{General Framework classes}
\begin{description}[leftmargin=5mm]
\item[{\WWClass}ScriptRunner:] entry point to the {\ww} simulation. It is responsible for building the environment necessary for a {\ww} script, executing the state machine, and setting up monitoring of the results.

\item [{\WWClass}Monitor:] responsible for recording events and collecting data for post-analysis, as the {\WWClass}Script is run.

\item [{\WWClass}Objective:] 
represents an objective that a {\WWClass}Script is aiming to achieve. 
Possible objectives include time, steps, episodes, `results' (such as vulnerabilities found, etc.).
The objective is also used to determine when to end the simulation.

\item [{\WWClass}Model:] a machine learning model for the bot, e.g., a Policy Gradient model with determined parameters.
\end{description}

\subsubsection{Per-use-case classes}
The core {\ww} platform consists of the general framework class together with a set of components from which the per-use case classes are defined.
In order to define each use case, we simply define a script and a bot class, using the components and deploy them on the general framework.

\begin{description}[leftmargin=5mm]
\item [{\WWClass}Script:]
describes the user community (e.g., the size of the graph),
and the environment where the users will interact (e.g., groups with fake news).

\item [{\WWClass}Bot:] 
an automated agent (represented by a test user) with a particular behaviour defined by actions.
For example, a FACEBOOK Messenger user.
A bot interacts with other users (as defined by its behaviour), and can have its own learning model.

\end{description}

\section{Applications of {\ww} at FACEBOOK}

We believe many of these application use cases for {\ww} may generalise to other WES systems, but we give them here in their FACEBOOK context to illustrate WES applications with specific concrete use cases.
At the time of writing we are focusing our engineering effort on the applications of {\ww} to integrity challenges, but we fully anticipate application to the other areas listed in the section in due course.
Indeed, we expect many more use cases to emerge as the development of the {\ww} infrastructure matures.

\subsection{Integrity and Privacy}
In any large scale system, not all user behaviour will be benign; some behaviours are perpetrated by bad actors, intent on exploiting the platform and/or its users.
On the Facebook platform such bad actor user behaviour includes any actions that contravene FACEBOOK's Community Standards~\cite{community-standards}.

We are using {\ww} to enhance our ability to detect  bad actor behaviours.
We are also developing  Automated Mechanism Design approaches that search product design space to find ways to harden the platform against such bad actors, 
thereby promoting the integrity of the platform and its users.
In this section, we illustrate both with  applications of {\ww} to the challenges of detecting and preventing contravention of integrity constraints.

{\ww} also provides us with a realistic, yet safely isolated, way to investigate potential privacy issues on the real platform. 
Because the {\ww} bots are isolated from affecting real users, they can be trained to perform potentially privacy--violating actions on each other.

On the other hand, because the bots use the real infrastructure, the actionability of any potential issues found through such simulation is considerably increased (compared to a traditional simulation on a purely theoretical model).

\noindent \textbf{Simulating bad actors}: Consider the problem of users trying to post violating content on the Facebook platform. 
Even though we employ state-of-the-art classifiers and algorithms to protect the platform, 
we need to be proactive in our exploration of the space of potential exploits; {\ww} provides one way to do this.
If our bots succeed in finding novel undetected contravening behaviours, the {\ww} simulation has allowed us to head off a potential integrity  vulnerability.

\noindent  \textbf{Search for bad content}: Bad actors use our platform to try to find  policy--violating content, or to try to locate and communicate with users who may share their bad intent.
Our bots  can be used to simulate such  bad actors,  exploring the social graph.
This enables them, for example,  to seek out policy--violating content and the violators that create it.
{\ww} bots can also search for clusters of users sharing policy--violating content.

\noindent  \textbf{Searching for mechanisms that impede bad actors}: 
We  use automated mechanism design to search for ways to frustrate these bad actors from achieving their goals within the simulation. 
This makes the optimisation problem a multi--objective one, in which the bots seek to achieve bad actions, while the system simultaneously  searches for mechanisms that frustrate their ability to do so. 
The search is also multi objective because we require new mechanisms that simultaneously frustrate the achievement of bad actions, while having little or no noticeable impact on normal users.
In this way we use {\ww} automated mechanism design to explore the space of potential changes that may also lead to greater integrity of the real platform.

Interestingly, this is a problem where {\em preventing} bad activity does not require the ability  to {\em detect} it.
Automated search may yield mechanisms that frustrate scamming, for example by hiding potential victims from scammers, without necessarily relying on classifiers that detect such scammers.
This is reminiscent of the way in which {\em removing} side effects (which may be computable),  does not require the ability to {\em detect} side effects (which is undecidable, in general) \cite{mhetal:icsm02}.

\noindent  \textbf{Bots that break privacy rules}: In the Facebook infrastructure, every entity has well--defined privacy rules. 
Creating bots trained to seek to achieve the sole purpose of breaking these privacy rules (e.g., to access another bot's private photos) is thus a way to surface potential bugs, as well as unexpected changes in the system's behaviour. 
For example, if a bot was never previously  able to perform a certain action (e.g., access another bot's message), but becomes able to do so after a code change,  this could highlight a change in privacy rules that resulted in unexpected behaviours.

\noindent  \textbf{Data acquiring bots}: Even with the privacy rules currently in place, a Facebook user has the ability to access another users' data (with consent, of course). 
This functionality is a necessary part of the normal usability of the platform.
Nevertheless, we need to maintain a constantly vigilant posture against any potential to exploit this fundamentally important ability. 
By creating bots whose sole purpose is to accrue as  much data as possible from each other, we are able to test our preventative measures and their effectiveness against this type of behaviour.

\subsection{Reliability}

Large organisation like Facebook naturally face challenges of reliability at scale.
{\ww} is not only a simulation of hundreds of millions of lines of code; it is a software system that runs on top of those very same lines of code.
In order to cater for the reliability of the {\ww} system itself, we use a DevOps approach, commonly used throughout the company~\cite{feitelson:deployment}.
{\ww} runs in continuous deployment as a production version, underpinned by suitable maintenance procedures, such as time series monitoring and analysis, alarms and warnings and on-call rotations.
However, {\ww} can also be used to explore the reliability of the platform as we outline in this section.

\begin{figure}
\centering
    \includegraphics[width=60mm]{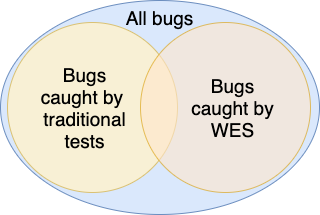}
    \caption{Traditional testing methods such as unit tests, or end-to-end tests, only succeed in capturing a subset of all possible bugs. {\ww} gives developers the option to test on a new level of abstraction, one that takes community behaviour and interactions into account.}
\end{figure}

\noindent  {\bf Social Bugs}: 
{\ww}  provides tools for social testing, whereby failures can be expressed as {\em percentages}, rather than more traditional binary success/fail. 
All traditional tests might execute successfully, yet we still observe an `social bug' issue in production. 
Examples include 
drops in top line product metrics, 
significant changes in machine learning classification outcomes,
big jumps in data pipeline line breakages.

These kind of bugs have many causes including code, data and/or configuration changes. 
While all could, {\em in theory} be detected by lower levels of test abstraction, it is useful
to have a WES style final `full ecosystem' test (as opposed to 
`full system' test). 
With
{\ww} we can detect such a significant  metric change before such a  change affects real users, because it tests the whole ecosystem with a simulation of the community that uses the platform.
A {\em single user} test, even at full system level, would be insufficiently expressive to capture community interaction faults.

We also retain lower levels of testing abstraction.
The {\ww} simulation is the most computationally expensive form of testing we have, as well as the highest level of abstraction.
Also, although `platform realism' is the goal of all WES systems, there are necessary compromises to achieve scalability, as discussed in Section~\ref{sec:open}.

\noindent  {\bf The WES Test Oracle}:   
These metrics play the role of test oracle~\cite{ebetal:oracle}, thereby ensuring that the platform level testing problem can be entirely automated.
Of course, since {\ww} is a scaled down version of the real community, there is a need to tune such metrics and alerts, but this is, itself, an interesting machine learning problem. 

\section{Open Problems and Some Related work that May Help Tackle Them}
\label{sec:open}

In this section we review related work and highlight  open problems and challenges for the WES research agenda.
Neither our characterisation of related work, nor our list of open problems is comprehensive.
We are excited to work with the academic and scientific research community to tackle  these open problems together using such related work and/or other promising approaches.

Naturally, we can expect  research to  draw on the extensive body of work on simulation, and in particular, multi agent simulation~\cite{michel:multi}, which is now relatively mature, with the advent of so-called `sophisbots' that are hard to distinguish from real users~\cite{boneh:relevant}.

There are also frameworks for simulation of software systems and communities, but these tend to focus on traditional simulation rather than on-platform simulation, the {\it sine qua non} of a WES system.
For example,
RecSym~\cite{Ie:RecSym} uses an abstraction of a generic Recommender System (RS) infrastructure to simulate the behaviour of different choices of Reinforcement Learning (RL) for recommending content to users. 
The most important difference, is that {\ww} uses RL (and other techniques) to train bots to behave like users so that the behaviours of users on the real Facebook infrastructure can be better simulated, whereas RecSym simulates the behaviour of an abstraction on Infra with respect to a given RL.  

\subsection{Open Problems and Challenges}
Since WES systems, more generally, rely on training bots to simulate real users on real software platforms, there is a pressing need for further research on a number of related topics. 
This section lists 15 areas of related work that can contribute to tackling open WES research agenda challenges.
The large number and diversity of  topics and challenges underscores the rich opportunities for research.

\noindent  {\bf Another Application for MARL}:
Recent developments in Multi Agent Reinforcement Learning (MARL)~\cite{jennings:agents} may offer techniques that can be adapted and applied to WES systems. 
One important challenge is to find ways to train bots to behave like specific classes of bad actors.

\noindent  {\bf Multi Objective Search}: 
Typically, Software Engineering problems, such as  reliability and integrity, will have a multi objective character. 
For example, it would be insufficient to constrain a mechanism to frustrate bad actors, 
if one does not counter-balance this objective against the (potentially competing) objective of not impeding normal users in their routine use of the platform.
Fortunately, multi objective optimisation algorithms, such as NSGA II~\cite{deb:nsga2}, are readily available  and have been widely--studied in the Software Engineering community for over two decades~\cite{mhbj:manifesto}.
More research is needed on the application of multi objective search to WES problems.

\noindent  {\bf AI Assisted Game Play}:
The WES  agenda naturally draws on previous work on artificial intelligence for software game play. 
Recent advances on competitive game playing behaviour~\cite{oriol:alphastar} 
may be adapted to also imbue WES  bots with more realistic social interactions. 
In a WES system we do not necessarily need competitive `play', but realistic social interaction; the rewards and fitness functions may differ, but key insights may, nevertheless, carry over between these two related application paradigms.

In some WES applications it may also be important to avoid the bots acquiring super-human abilities, such as interacting faster than any human ever could.
This is also a problem recently tackled in machine learning for computer game playing optimisation~\cite{oriol:alphastar}.

\noindent  {\bf Automated Mechanism Design}: 
 
Mechanism Design is a form of automated game design, a topic that has been studied for over a decade~\cite{togelius:experiment}, and that remains an active area of research  ~\cite{kunanusont:n-tuple}. 
The challenge is to define techniques for deployment, testing and for efficiently and effectively searching the mechanism space. 
Tackling these problems may draw on previous work on 
Genetic Improvement~\cite{Petke:gisurvey}, 
Program Synthesis~\cite{gulwani:program-synthesis}, 
Constraint Solving~\cite{kang:constraint}, 
and 
Model Checking~\cite{clarke:model-checking}.

\noindent  {\bf Co-evolutionary Mechanism Learning}:
Automatically improving the platform mechanism to frustrate some well-known attack from a class of bad actors may yield short term relief from such bad actors.
However, in order to continue to remain `ahead of the game' and to thereby frustrate  all possible, as yet unknown attacks, we need a form of 
co-evolutionary optimisation; the mechanism is optimised to frustrate bad actions, while the bots simultaneously learn counter strategies that allow them to achieve these bad actions {\em despite} the improved mechanism.

Co-evolutionary optimisation is well-suited to this form of `arms race'.
Co-evolutionary optimisation has not yet received widespread attention from the SBSE community, 
although there is some initial literature on the topic~\cite{Arcuri2010,jretal:ssbse11,kostas:gecco04}.
Co-evolutionary Mechanism Design therefore establishes an new avenue of research that promises to widen the 
appeal and application of co-evolutionary approaches to software engineering problems.

\noindent  {\bf End User Realism and Isolation}:
In some applications, WES bots will need to be trained to faithfully model the behaviours of the platform's real users; `end user realism'. 
Tackling this may rely on recent advances in machine learning, 
but will also be constrained by the need for user privacy. 
There is also interesting research to be done on the degree of end user realism required, and metrics for measuring such realism, a problem only hitherto lightly explored in testing research~\cite{afshan:evolving,bozkurt:automatically,draheim:realistic,ke:ieee-se17}. 

Because bots are isolated from real users, we face the research challenge of 
defining,
capturing, 
measuring,
and
replicating 
realistic behaviour.
Faithfully replicating every aspect of end user behaviour is seldom necessary. 
Indeed, in some cases, end user realism is not required at all.
For example, for social testing the FACEBOOK Messenger system, 
we found that it was  sufficient to have a community of bots regularly sending messages to one another in order to detect some social bugs that manifest through drops in production metrics, such as number of messages sent.

For integrity-facing applications, such as preventing bad actors' harmful interaction with normal users, we need reasonably faithful bad actor bot behaviours, and bots that faithfully replicate normal users' responses to such bad actors.
This is a challenging, but highly impactful research area.

\noindent  {\bf Search Based Software Engineering}:
Many of the applications of WES approaches lie within the remit of software engineering, and it can be expected that software engineering, and in particular, Search Based Software Engineering (SBSE)~\cite{mhbj:manifesto} may also find application to open problems and challenges for WES systems. 
In common with SBSE, WES systems share the important salient characteristic that the simulation is executed on the real system itself.
It is therefore `direct'.
This directness is one advantage of SBSE over  other engineering applications of computational search ~\cite{mh:fase10-keynote}. 
We can expect similar advantages for WES systems.
By contrast, traditional simulation tends to be
highly indirect:
The simulation is not built from the real system's components, but as an abstraction of a theoretical model of the real system and its environment.

\noindent  {\bf Diff Batching}: 
The WES approach has the advantage that it  allows engineers to investigate properties of proposed changes to the platform. 
However, for larger organisations, the volume of changes poses challenges itself. 
For example, at FACEBOOK, over 100,000 modifications to the repository land in master every week~\cite{mhetal:ssbse18-keynote}.
Faced with this volume of changes, many companies, not just FACEBOOK \cite{najafi:bisecting}, use Diff batching, with which collections of code changes are clustered together.
More work is needed on smarter clustering techniques that group code modifications (aka Diffs) in ways that promote subsequent bisection \cite{najafi:bisecting}.

\noindent  {\bf Speed up}:
Simulated clock time is a property under experimental control.
It can be artificially sped up, thereby yielding results in orders of magnitude less real time than a production A/B test~\cite{siroker:ABTesting}. 
However, since a WES system uses real infrastructure,  we cannot always speed up behaviour without introducing non-determinism: 
If bots interact faster (with the system and each other)  this may introduce race conditions and other behaviours that would tend to be thought of as flakiness in the testing paradigm~\cite{luo:flaky,mhpoh:scam18-keynote}.

\noindent  {\bf Social Testing}:
Section~\ref{sec:st} introduces a new form of software  testing, which we call `Social Testing'. 
Testing is generally regarded as an activity that takes place at different levels of abstraction, with unit testing typically being regarded as lowest level, while system level testing is regarded as highest level.
Social testing adds a new level of abstraction above system level testing.
There are so many interesting problems in social testing that a complete treatment would require a full paper in its own right. 
In this brief paper we hope we have sufficiently motivated the introduction of this
new higher level of abstraction, and that others will be encouraged to take up research on social testing.

\noindent  {\bf Predictive Systems}: 
WES systems would benefit from automated  prediction (based on the simulation) of the future properties of the real world.
This will help translate insights from the simulation to actionable implications for the real world phenomena.
Therefore, research on predictive modelling~\cite{catal:systematic,mh:promise10-keynote} is
also highly relevant to the WES research agenda.

\noindent  {\bf Causality}: 
To be actionable,  changes proposed  will also need to correlate with improvements in the real world, drawing potentially on advances in causal reasoning~\cite{pearl:causality}, another topic of recent interest in the software engineering community~\cite{wmetal:causal}.

\noindent  {\bf Simulating Developer Communities}: 
Although this paper has focused on WES  for social media users, a possible avenue for other WES systems lies in simulation of developer communities. 
This is a potential new avenue for the  Mining Software Repositories (MSR) research community~\cite{hassan:road}. 
The challenge is to mine information that can be usefully employed to train bots to behave like developers, thereby exploring emergent developer community properties using WES approaches. 
This may have applications to and benefit from MSR. 
It is also related to topics such as App Store analysis~\cite{martin:tse-survey}, for which the community combines developers and users, 
and to software ecosystems ~\cite{manikas:ecosystems-slr},
which combine diverse developer sub-communities.

\noindent  {\bf Synthetic Graph Generation}:
For {\ww}, we are concerned with the simulation of social media.
Read-only bots can operate on the real social network, which is protected by isolation.
However, many applications require writer bots. 
Naturally, we do not want {\ww} writer bots interacting with real users in unexpected ways, so part of our isolation strategy involves large scale generation of large synthetic (but representative) graphs. 
This is an interesting research problem in its own right.
On a synthetic graph it will also be possible to deploy fully isolated bots that can exhibit arbitrary actions and observations, without the need for extra mechanism constraints to enforce isolation.

\noindent  {\bf Game Theory}:
A WES execution is a form of game, in which both the players and the rules of the game can be optimised (possibly in a co-evolutionary `arms race').
Formal game theoretic investigation of simulations offers the possibility of underpinning the empirical observations with mathematical analysis. 
Naturally, empirical game-theoretic analysis \cite{wellman:empirical} is also highly relevant.
There has been recent interest in game theoretic formulations in the Software Engineering community \cite{carlos:assessor}.
WES systems may provide a further stimulus for this Game Theoretic  Software Engineering research agenda.

\section{Conclusion}

In this paper we set out a new research agenda: {\wwbacronym} of user communities.
This WES  agenda draws on
 rich research strands,   including machine learning and optimisation, multi agent technologies, reliability, integrity, privacy and security as well as traditional simulation, and topics in user community and emergent behaviour analysis.   
 
 The promise of WES  is realistic, actionable, on-platform simulation of complex community interactions that can be used to better understand and automatically improve deployments of multi-user systems.
In this short paper, we merely outline the WES research agenda and some of its open problems and research challenges.
Much more remains to be done.
We hope that this paper will encourage further uptake and research on this exciting WES research agenda.

\section{Acknowledgements}
The authors would like to thank Facebook's engineering leadership for supporting this work and also the many Facebook engineers who have provided valuable feedback, suggestions and use cases for the FACEBOOK {\ww} system.

\balance
\bibliography{slice}
\bibliographystyle{ACM-Reference-Format}
\vspace{12pt}
\end{document}